\documentclass{elsart}
\input psfig.sty
\usepackage{natbib}

\newcommand{\oiii}{{\sc O\,iii}}

\newcommand{\hi}{{\sc H\,i}}
\newcommand{\feii}{Fe\,{\sc ii}}

\begin{document}
\runauthor{Pogge}

\begin{frontmatter}
\title{Narrow-Line Seyfert 1s: 15 Years Later}
\author[OSU]{Richard W. Pogge}
\address[OSU]{Department of Astronomy, The Ohio State University, 
        Columbus, OH 43210-1173, USA}

\begin{abstract}
The spectroscopic properties of the objects that came to be called
Narrow-Line Seyfert 1s were first systematically described and named as
such 15 years ago by Osterbrock \& Pogge (1985).  At the time, they were a
relatively rare and peculiar subclass of Seyfert galaxies.  Their discovery
in large numbers in X-ray surveys, however, has elevated them to a role as
important members of the AGN family, ones which may hold many keys to
understanding the physics of AGN across the electromagnetic spectrum.
This contribution reviews the spectral classification of the narrow-line
Seyfert 1s, and describes some of the properties that make this unusual
class of objects interesting.
\end{abstract}
\end{frontmatter}

\section{Narrow-Line Seyfert 1s}

\begin{quote}
{\it ``This unusual object merits further observations...''}
\begin{flushright}
Davidson \& Kinman (1978)\\
{\it On the possible importance of Markarian 359}
\end{flushright}
\end{quote}

The discovery of a new class or subclass of astronomical objects almost
never begins with the ``definitive'' paper, since that paper is more often
than not the outcome of a more systematic investigation into previous
reports that some objects might be interesting.  So too with narrow-line
Seyfert 1s (NLS1s).  

The watershed year in the prehistory of NLS1s is 1978.  Davidson \& Kinman
\cite{DK78} described the unusual and potentially important spectral
properties of Markarian 359.  Mrk\,359 has a Seyfert 1-like spectrum with
unusually narrow permitted lines; the FWHM of H$\beta$ was
520$\pm$100~km~s$^{-1}$, comparable to what is seen in classic Seyfert 2s.
The second curiosity of 1978 was Mrk\,42.  Koski \cite{K78} and Phillips
\cite{P78} remarked that Markarian 42 had many of the spectral properties
of Seyfert 1s, but that the \feii\ and \hi\ lines were very narrow, like in
Seyfert 2s.

In 1983, Osterbrock \& Dahari \cite{OD83} undertook a systematic
classification of a sample of Seyferts and candidate Seyferts, and noted in
their tables that four showed unusual properties, namely ``narrow \hi\ and
\feii\, ...\feii\ strong'' (Mrk 493), and ``Very narrow \hi\ ... but
noticeably wider than [\oiii]'' (Mrk 783).  It was work that led Don
Osterbrock to undertake a systematic search for and study of those objects
which, along the lines first noted by Davidson \& Kinman, had all the basic
spectral characteristics of Seyfert 1s but unusually narrow permitted
lines.  It was this project that Don invited me to join as a new graduate
student at UC Santa Cruz in 1984.  The rest, as they say, is history...

In Osterbrock \& Pogge \cite{OP85}, we defined NLS1s to be those galaxies
whose nuclear spectra are generally like those of Seyfert 1s (strong \feii,
[\oiii] relatively weak compared to the Balmer lines), but with line widths
much narrower than typical Seyfert 1s.  The formal spectral classification
criteria for NLS1 galaxies that has emerged since are:
\begin{itemize}
\item Narrow permitted lines only slightly broader than the forbidden
      lines.
\item {[\oiii]/H$\beta < 3$}, but exceptions are allowed if there is
      also strong [Fe\,{\sc vii}] and [Fe\,{\sc x}] present, unlike
      what is seen in Seyfert 2s.
\item FWHM(H$\beta$)$<$2000 km~s$^{-1}$.
\end{itemize}
The first two criteria are from our original classification \cite{OP85},
while the maximum line-width criterion was introduced by Goodrich
\cite{G89} in his spectropolarimetric study.

\begin{figure}
\psfig{figure=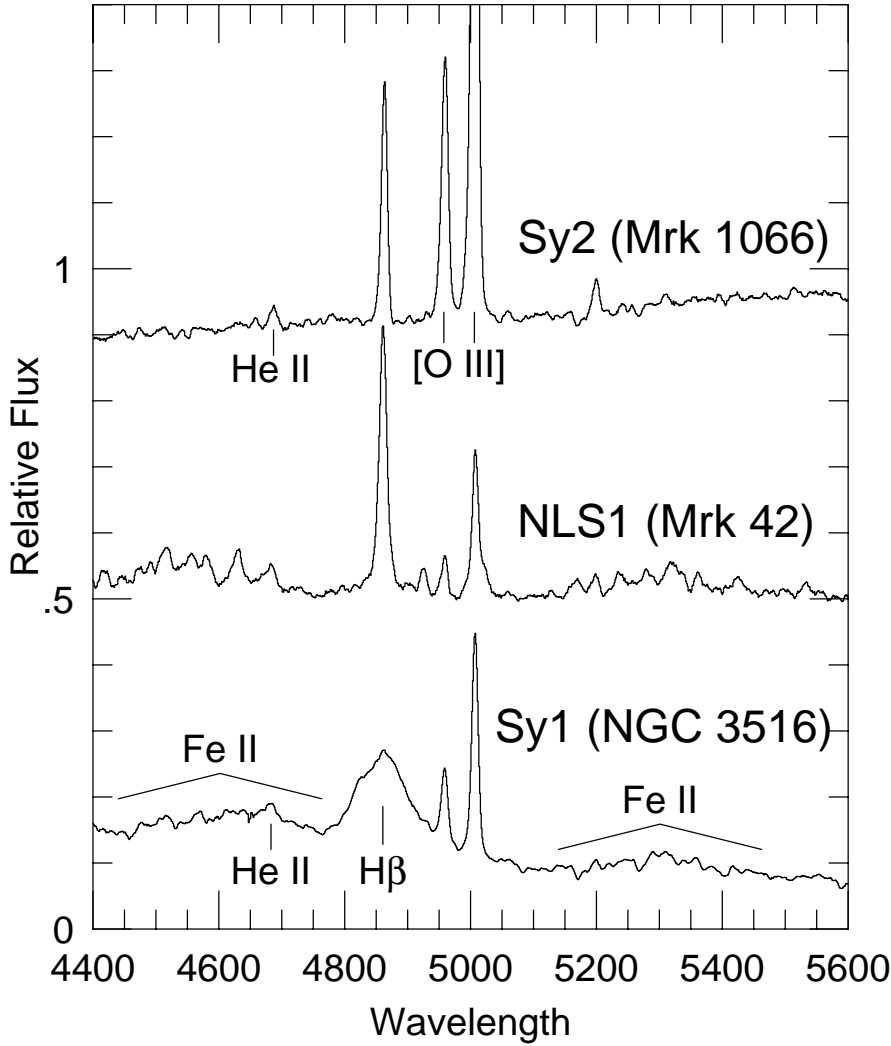,width=12.0cm}
\caption{Spectra in the region of H$\beta$ of the NLS1 Mrk\,42 (center),
the Sy1 NGC\,3516 (below), and the Sy2 Mrk\,1066 (above).}
\label{fig:fig1}
\end{figure}

The essence of the classification is shown graphically in
Figure~\ref{fig:fig1}, which contrasts an NLS1 against Seyfert 1s and 2s.
While the H$\beta$ line of the NLS1 Mrk\,42 is not much wider than the
Seyfert 2 Mrk\,1066, the other spectral lines, especially [\oiii] and
\feii, appear in about the same proportions, if with narrower \feii\
widths, as they do in Seyfert 1s like NGC\,3516.

\section{I\,Zw\,1 and the Elusive Type 2 QSOs}

\begin{quote}
{\it ``So, What is 1E0449.4$-$2834 Anyway?''}
\begin{flushright}
Halpern, Eracleous, \& Forster (1998)
\end{flushright}
\end{quote}

NLS1s appear to have analogs among the more luminous QSOs, so-called
narrow-line QSOs.  The prototype of the narrow-line QSOs is arguably
I\,Zw\,1 \cite{L97}.  The luminosity of I\,Zw\,1 places it at the ragged
boundary between the most luminous Seyfert 1s and the least-luminous QSOs,
an arbitrary dividing line set at $M_B=-23$\,mag \cite{SG83}.  Sargent
\cite{S68} had noted as early as 1968 that I\,Zw\,1 had unusually strong
\feii\ emission.  Phillips \cite{P76, P77} made landmark studies of the
\feii\ lines in this object, assisted both by their strength and by the
fact that the lines were sufficiently narrow that you could definitively
identify the components of line blends that are otherwise just
undifferentiated emission ``humps'' in normal Seyfert 1s or QSOs.  The
unusually narrow lines of I\,Zw\,1's were not lost on others, for example
Oke \& Lauer \cite{OL79} who in 1979 commented that ``I\,Zw\,1 is not a
typical type 1 Seyfert since the permitted and forbidden lines are of
comparable breadth''.

The first connection between NLS1s and objects like I\,Zw\,1 was made by
Halpern \& Oke \cite{HO87} in their 1987 study of Mrk\,507 and 5C3.100.
They were drawn to these objects by their unusually large X-ray
luminosities compared to other Seyfert 2s.  Their superior spectra showed
that these objects were, in fact, not Seyfert 2s, but rather had spectra
like I\,Zw\,1 (strong \feii\ and narrow permitted lines).  This made them
NLS1s with luminosities near the high-end of the range for Seyfert 1s.
This paper is also notable in that they are the first to call attention to
the possible importance of NLS1s as X-ray sources.

The story of the original Seyfert 2 classification of Mrk\,507 and 5C3.100
is instructive.  The earlier spectra showed the strong narrow H$\beta$ and
[\oiii] lines, but the \feii\ lines were lost in the noise; hence they were
thought to be Seyfert 2s.  Halpern \& Oke's better spectra showed the
\feii\ lines clearly, removing their Seyfert 2 assignment.  Similarly,
Schmidt \& Green \cite{SG83} rejected Mrk\,684 from their list of bright
QSOs in 1983 because it had ``narrow emission lines, such as those observed
in Seyfert 2 galaxies''.  Don and I re-observed this object and noted
\cite{OP85} that in fact Mrk\,684 was an NLS1 with ``strong \feii\ emission
and weak or nonexistent forbidden lines.''  The false classification of
objects as Seyfert 2s because of poor signal-to-noise ratio masking either
weak \feii\ or weak broad lines (or both) has been a persistent problem.
Nowhere is this kind of classification confusion more chronic than in the
long and so far fruitless search for QSO analogs of the Type 2 Seyferts;
objects with QSO luminosities but Seyfert 2-like spectra (i.e., no broad
lines).  At issue is whether the full range of spectral properties found
among lower luminosity AGN like Seyferts is recapitulated at higher
luminosities.

Every now and then a preprint or paper comes to my attention that declares
that such and such an object is the ``first'' Type 2 QSO.  For a long time
I would dutifully write down its name and citation on an unused corner of
my office blackboard and wait.  Sure enough, eventually a paper by Jules
Halpern and his collaborators would come along presenting a deeper, lower
noise spectrum and reclassifying the object as either an NLS1 (e.g., IRAS
20181$-$224 \cite{ES94, HM98}, AXJ0341.4$-$4453 \cite{S82, HEF98}) or an
intermediate Seyfert galaxy with weak broad lines (e.g., 1E0449.4$-$1823
\cite{B98, HTG99}).  I called it my ``Jules Just Says No'' list.  So far,
Jules and collaborators are ahead.

Why is this?  On the technical side is the inescapable fact of
spectrophotometry that signal-to-noise ratio matters.  Poor spectra give
poor (or even wrong) classifications.  Further, Seyfert 2s are not just
characterized by narrow emission lines, but also by a conspicuous {\it
absence} of weak broad lines or \feii\ emission.  Detection of these
requires good spectra; the characteristic narrow lines are the easy part of
the observation.  On the scientific side is the possibility that true Type
2 QSOs are extremely rare, so rare that after nearly 4 decades of AGN
research there is not yet a definitive example that hasn't eventually been
reclassified with better spectroscopy.  The operative word here is ``yet'',
since the search continues regardless; rarity has its attractions, and
these appear to be among the rarest of the rare.  So far what it tells us
is that the differences between Seyferts and QSOs are not simply a matter
of luminosity.

This brings up the issue of how to classify AGN.  The division of AGN into
Type 1 and Type 2 is based on well-defined and well-established spectral
criteria.  While we hope that our classification has the virtue of relating
to the underlying physics of AGN, which relation is still imperfectly
understood, it is still fundamentally empirical in nature.  This should be
obvious, but there has been a new fashion of late among some astronomers to
call heavily obscured QSOs ``Type 2 QSOs'', even though their spectra do
not have the characteristics of Seyfert 2s (i.e., narrow permitted and
forbidden lines, high-excitation line ratios, etc.).  A dusty QSO
is a dusty QSO, it is not a Type 2 QSO unless its optical
spectrum satisfies the spectral criteria of a Seyfert 2.

\section{X-Rays and the ROSAT Renaissance}

\begin{quote}
{\it ``X-ray selection may be an efficient way to find narrow-line\\
      Seyfert 1 galaxies.''}
\begin{flushright}
Stephens~(1989)
\end{flushright}
\end{quote}

Sally Stephens' PhD dissertation at UCSC was a spectroscopic study of 65
X-ray selected AGN.  Of these, 10 were NLS1s, or $\sim$15\%, leading to the
quote above taken from the abstract of her paper \cite{SS89}.  Subsequent
work by Puchnarewicz and collaborators found $\sim$50\% \cite{P92,P95} of
their ultra-soft X-ray selected AGN were NLS1s.  In hindsight, an
examination of the spectra of AGN found in the HEAO-1 survey of Remillard
et al. \cite{R86} and the Einstein MSS of Gioia et al. \cite{G84} found
roughly similar proportions of NLS1s among them, if they were unrecognized
as such at the time.

The first indication I had that Sally's comment was truly prophetic was at
the 1993 IAU Symposium 159 in Geneva.  There Dirk Grupe showed me his
poster paper \cite{G93} describing the results of a follow-up study of 40
new Seyferts discovered in the ROSAT All-Sky Survey.  Grupe et al. found
that $\sim$50\% of the RASS soft X-ray selected AGN were NLS1s.  In that
same year, Boller et al. \cite{B93} inaugurated the ``ROSAT Renaissance''
in the study of NLS1s with their seminal paper on the truly outlandish X-ray
variability of IRAS 13224$-$3809.  This NLS1 with strong \feii\ emission
increased in 0.1$-$2.4 keV brightness by a factor of 4 with an unheard of
(for Seyferts) doubling time of 800 seconds!  Not only are NLS1s common in
soft X-ray selected samples, they are among the most variable AGN known
outside of extreme objects like blazars.

In 1996 the ROSAT team effectively assumed ownership of the NLS1 class. In
their important paper of that year, Boller, Brandt, \& Fink \cite{BBF96}
published their results for a study of 46 NLS1s, about half of which were
X-ray {\it discovered}.  This paper demonstrated the remarkable soft X-ray
properties of NLS1s:
\begin{itemize}
\item Wide range of 0.1$-$2.4 keV photon indices, $\Gamma\approx1-5$, 
      compared to $\sim 2.1$ for typical Seyfert 1s.
\item Rapid, high-amplitude X-ray variability (doubling times of 
      minutes to hours).
\end{itemize}
While some NLS1s have the steepest soft X-ray excesses yet observed, it is
clear from Figure 8 of Boller, Brandt \& Fink \cite{BBF96} that not all
NLS1s are ultra-soft excess sources.  A number of spectroscopically
classified NLS1s have photon indices more typical of the general run of
Seyfert 1s.  A soft X-ray excess is a common, but not defining,
characteristic of NLS1s.  Searching for ultra-soft X-ray sources has proven
to be an excellent way to find new NLS1s, but it is biased against NLS1s
with harder 0.1$-$2.4\,keV X-ray spectra.  While the demographics of NLS1s
among AGN in general is still poorly understood, the sense I get is that
despite their abundance in soft X-ray surveys they are still relative
rarities in general (see papers by Grupe and Hasigner herein).

\section{Lies, Damned Lies, and Principal Components Analysis}

\begin{quote}
{\it ``They clearly demonstrate that the Seyfert phenomenon is not a 
     simple one-parameter effect.''}
\begin{flushright}
Osterbrock~\&~Pogge~(1985)
\end{flushright}
\end{quote}

Determining the role of NLS1s within the AGN phenomenon may be approached
observationally by understanding where they fall among the statistical
properties of the entire class.  In this regard NLS1s again display their
remarkable propensity for seeking out extremes.  In Boroson \& Green's
landmark 1992 study of the emission-line properties of low-redshift QSOs
\cite{BG92}, their Principal Components Analysis (PCA) revealed two
convincing eigenvectors among a variety of emission-line and continuum
measurements for 87 QSOs in the BQS catalog with $z<0.5$.  The second
eigenvector is essentially a relatively weak H$\beta$ Baldwin Effect.  The
principal eigenvector, the so-called Boroson \& Green Eigenvector 1, is
stronger but its underlying physical basis has proven more elusive.

The principal driver behind Eigenvector 1 is a strong anticorrelation
between the strengths of the \feii\ and [\oiii]$\lambda$5007 emission
lines.  Additional contributions come from a correlation between the FWHM
of H$\beta$ and the peak of [\oiii].  At one extreme end of AGN along
Eigenvector 1 are those objects with the strongest \feii, the weakest
[\oiii], and the narrowest H$\beta$ lines; all of which are the defining
spectral characteristics of NLS1s.

Statistically speaking, PCA is a blunt instrument, a mathematical {\it
bellum omnes contra omnium} among many different measurements in the hope
that the ultimate physical driver of the phenomenon under study will be
revealed in the principal eigenvector.  It is not yet clear to me that the
Boroson \& Green Eigenvector 1 has lived up to this expectation, but it
{\it is} interesting that NLS1s are all clustered at one extreme end of
this eigenvector.  What is this trying to tell us?  I don't know, but at
the very least the properties of the NLS1s forcefully demonstrate that AGN
are not a one-parameter family of objects.  NLS1s {\it look} superficially
like Seyfert 1s, but their many unusual and extreme properties are clearly
telling us that they are more than just Seyfert 1s with narrow lines.

My thanks to the workshop organizers, the WE-Heraus-Stiftung, and the
Phyzikzentrum Bad Honnef for a stimulating workshop.  Travel was supported
by the Ohio State University Department of Astronomy.  My sincerest thanks
to my teacher, mentor, dissertation advisor and scientific godfather
Prof. Donald Osterbrock, who made me an offer I couldn't refuse 15 years
ago and gave me my start in AGN.


\begin{thebibliography}{999}
\bibitem{B93}   Boller, Th., et al. 1993, A\&A, 279, 53 
\bibitem{B98}   Boyle, B.J. et al. 1998, MNRAS 297, L53.
\bibitem{BBF96} Boller, Th., Brandt, W.N., \& Fink, H. 1996, A\&A, 305, 53
\bibitem{BG92}  Boroson, T. \& Green, R. 1992, ApJS, 80, 109
\bibitem{DK78}  Davidson, M.K., \& Kinman, T.D. 1978, ApJ, 225, 776
\bibitem{ES94}  Elizade, F. \& Steiner, J.E. 1994, MNRAS, 268, L47
\bibitem{G84}   Gioia et al. 1984, ApJ, 283, 495
\bibitem{G89}   Goodrich, R.W. 1989, ApJ, ApJ, 342, 234
\bibitem{G93}   Grupe, D. et al. 1993, IAU Symposium 159.
\bibitem{HEF98} Halpern, J.P., Eracleous, M., \& Forster, K. 1998, ApJ, 501, 103
\bibitem{HM98}  Halpern, J.P. \& Moran, E.C. 1998, ApJ, 494, 194
\bibitem{HO87}  Halpern, J. \& Oke, J.B. 1987, ApJ, 312, 91
\bibitem{HTG99} Halpern, J.P. Turner, T.J., George, I.M. 1999, MNRAS, 307, L47
\bibitem{K78}   Koski, A. 1978, ApJ, 223, 56
\bibitem{L97}   Laor, Januzzi, Green, Boroson 1979, ApJ, 489, 656
\bibitem{OD83}  Osterbrock, D.E. \& Dahari, O. 1983, ApJ, 273, 478
\bibitem{OL79}  Oke, J.B. \& Lauer, T.R. 1979, ApJ, 230, 360
\bibitem{OP85}  Osterbrock, D.E. \& Pogge, R.W. 1985, ApJ, 297, 166
\bibitem{P76}   Phillips, M.M. 1976, ApJ, ApJ, 208, 37
\bibitem{P77}   Phillips, M.M. 1977, ApJ, ApJ, 215, 746
\bibitem{P78}   Phillips, M.M. 1978, ApJS, 38, 187
\bibitem{P92}   Puchnarewicz, E.M. et al. 1992, MNRAS, 256, 589
\bibitem{P95}   Puchnarewicz, E.M. et al. 1995, MNRAS, 276, 20
\bibitem{R86}   Remillard et al. 1986, ApJ, 301, 742
\bibitem{S68}   Sargent, W.L.W. 1968, ApJ, 152, L31.
\bibitem{S82}   Stocke, J. et al. 1982, ApJ, 252, 69
\bibitem{SG83}  Schmidt, M. \& Green, R. 1983, ApJ, 269, 352
\bibitem{SS89}  Stephens, S. 1989, AJ, 97, 10
\end{thebibliography}
\end{document}